# Haptic Sensing for MEMS with Application for Cantilever and Casimir Effect

M. Calis[1,2], M.P.Y Desmulliez[2]

[1]Corac Group plc
Brunel Science Park
Uxbridge, UB8 3PQ

[2]MIcroSystems Engineering Center (MISEC)
School of Engineering & Physical Sciences
Heriot-Watt University, Edinburgh, EH14 4AS This paper presents an implementation of the Cosserat theory into haptic sensing technologies for real-time simulation of microstructures. Cosserat theory is chosen instead of the classical theory of elasticity for a better representation of stress, especially in the nonlinear regime. The use of Cosserat theory leads to a reduction of the complexity of the modelling and thus increases its capability for real time simulation which is indispensable for haptic technologies. The incorporation of Cosserat theory into haptic sensing technology enables the designer to simulate in real-time the components in a virtual reality environment (VRE) which can enable virtual manufacturing and prototyping. The software tool created as a result of this methodology demonstrates the feasibility of the proposed model. As test demonstrators, a cantilever microbeam and microbridge undergoing bending in VRE are presented.

## I. INTRODUCTION

Commercial MEMS are usually the result of many manufacturing, characterisation, packaging and tests iteration runs used to optimise their performance and reliability. At the design and manufacturing phases, engineers employ a variety of software tools that deal with the analysis of complex geometrical structures and the assessment of various component interactions that often belong to different domains of energy. Software packages such as CoventorWare [1], MEMSCAP [2] and ANSYS [3] are based however on classical elasticity, which limits the validity of stress results for large deflections of structures. In addition, these software packages tend to be computationally intensive when dealing with complex microsystems. Commercial software packages such as CoventorWare [1], MEMS Pro [4], SABER [5], IntelliSuite [6] do not allow real-time simulation. An efficient and rapid modelling approach that represents accurately the linear and nonlinear dynamic behaviors of MEMS is therefore called for.

## II. HAPTIC SENSING TECHNOLOGY

The term haptic originates from the Greek "haptesthai" and means the sense of touch with both *tactile* (cutaneous) and *kinesthetic* (proprioception) feedback. The PHANTOM Omni [7], an analogue interface, from the company "SensAble Technologies" enables the interaction of the user with virtual environments based on haptic rendering through touch. This haptic device is a "pen" (also called stylus, proxy and haptic interface point) located at the end of an arm controlled by electric motors. These new interfaces differ from current computer systems and are changing the way by which digital information is perceived and manipulated. One of the most powerful ideas behind haptic technologies is to design ideas creatively and not mathematically. A haptic system consists of two loops, called haptic loop and display loop that need to be maintained with an update rate of 1 KHz and 30 Hz, respectively (Figure 1) [8]. The display loop must be updated at 30 Hz since the Human Visual System (HVS) has a flicker fusion frequency around 30-60 Hz. The update rate of the haptic loop has been set at 1 KHz to avoid force artefacts and due to the human being able to detect discrete events at less than 1 KHz.

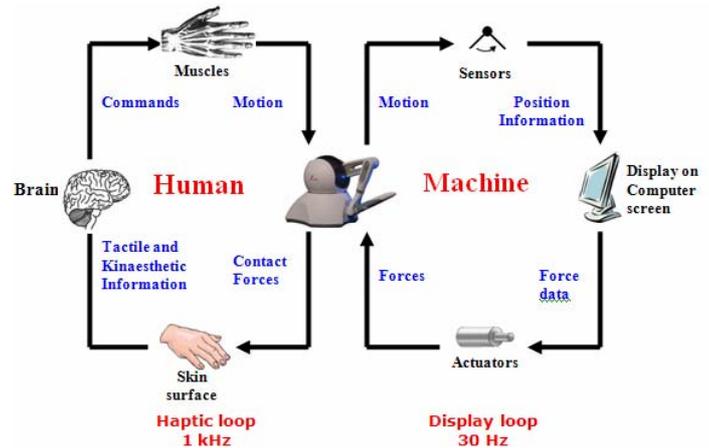

Fig. 1. Schematic of a haptic system

Haptic computers represent an important advantage in simulated industrial product development, virtual prototyping, in assembly/disassembly process, in final design and finally for training. Haptic computers enable also designers the possibility to scale-up to a human ergonomic scale micro-assemblies (such as MEMS) which can consist of millions of parts and then scale the product back to its original size. These properties improve the manufacturing lead times and the maintenance. In addition, in VREs, scaling down MEMS allow designers to gain more information and better understanding of their devices/components since the designers will be immersed in the VRE and any part of the system will be accessible. The different needs of MEMS modelling and simulation methodology for the precise performance verification of MEMS products are summarized as shown in table 1.

| Application | Needs | Benefits of haptic technologies |
|---|---|---|
| **CAD/CAM design** | • Guide designers during assembly and disassembly process<br>• Conception<br>• Tolerance | • Manufacturing of mould at micro-scales for LIGA processes<br>• Sense surface, shape of components, deformation<br>• Sense and effect of forces at micro-, meso- and nano-meters size structures |
| **Virtual Prototyping** | • Replace physical prototype by virtual model<br>• Enhance product development | • For combining physical and digital modeling<br>• Can guide designers<br>• Improve manufacturing lead times |
| **Visualization** | • Analyses of any parts of the system<br>• Ergonomic analysis | • Scale-up/down<br>• Increased information flow between user and the computer<br>• Better understanding |
| **Maintenance** | • Verification<br>• Diagnosis | • Quick analysis of any default, the cause and solution<br>• Security |
| **Training** | • Simulation<br>• A better understanding<br>• Useful for application related to manipulation | • Force-feedback<br>• Sense gravity, inertia<br>• Motion of components |

## III. COSSERAT THEORY

Methods for modelling MEMS components can usually be classified into two categories, as shown in Figure 2. The exact methods include the Euler-Bernoulli [9] and the Timoshenko techniques [10] which are solved using a power series expansion. The FEA, BEM and lumped mass methods are classified under the approximate methods and are solved using superposition techniques. The FEM is also known as the matrix displacement method. The word approximate is used since it assumes that displacements can be represented by simple polynomial expressions. Our approach uses a semi-analytical method based on both power series expansions and a multimodal approximate method [11].

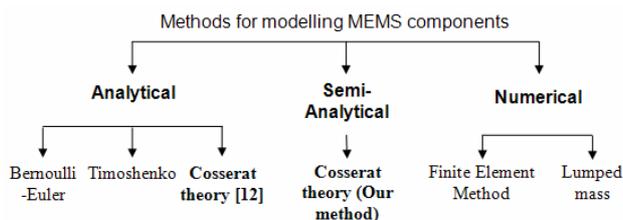

Fig. 2. Taxonomy of MEMS modelling methods

One reason of using the Cosserat theory is its capability to meet the update rate of both loops. To the best of the authors' knowledge, we are the first to implement Cosserat theory in haptic sensing technologies [8,12]. The motion in space of a nonlinear Cosserat rod segment can be represented as a vector $r(s,t)$, called a Cosserat curve, which describes the position of the line of centroids of the cross-sections (Figure 3, dotted line). Each $(s,t)$ is a right-handed orthonormal basis where $s$ denotes the length parameter of the rod segment ($a \leq s \leq b$), $t$ denotes the Newtonian time. $d_1(s,t)$ and $d_2(s,t)$ are a pair of orthogonal material lines describing the principal orientations and giving information about the location of the material cross section. $d_3$ is the normal to the cross-section and defined by

$$\mathbf{d}_3(s) = \mathbf{d}_1(s) \times \mathbf{d}_2(s) \qquad (1)$$

which provides an orthonormal frame $D = \{\mathbf{d}_k(s,t)\}$ with k =1, 2, 3. $\mathbf{d}_3(s)$ encodes the state of deflection of the beam at each point along this line.

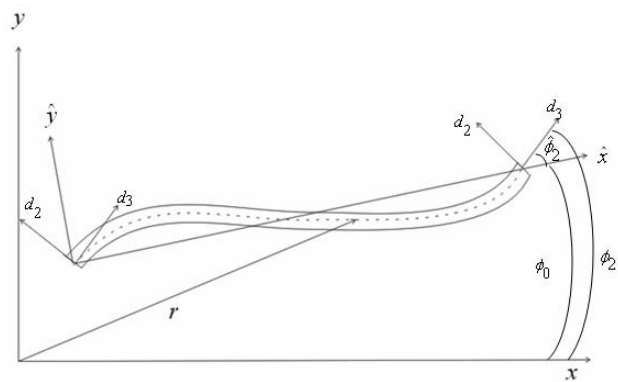

Fig. 3. A Cosserat rod: Deformed beam

The deformation of the slender MEMS structure represented by the deformation of the centroid line depends upon three vectors $\mathbf{r}(s)$, $\mathbf{d}_1(s)$ and $\mathbf{d}_2(s)$. In the Cosserat theory, the accuracy will depend on the method used to model the motion/deformation of the centroid line. Unlike in [12] where the Newton's dynamical law and analytical solution are used, our approach is based on a semi-analytical method and on the Euler-Bernoulli equation of motion. In the solution process, the displacements $u_x$ and $u_y$, in the transverse and axial directions, are expanded in ascending powers of $w$ [13].

$$\mathbf{u} = \begin{bmatrix} u_x \\ u_y \end{bmatrix} = \left( \begin{bmatrix} \mathbf{a}_{0x} \\ \mathbf{a}_{0y} \end{bmatrix} + w \begin{bmatrix} \mathbf{a}_{1x} \\ \mathbf{a}_{1y} \end{bmatrix} + w^2 \begin{bmatrix} \mathbf{a}_{2x} \\ \mathbf{a}_{2y} \end{bmatrix} + w^3 \begin{bmatrix} \mathbf{a}_{3x} \\ \mathbf{a}_{3y} \end{bmatrix} + .... \right) \qquad (2)$$

By rewriting (2) as

$$c^4 \sum_{r=0}^{\infty} w^r a_{ry}^{iv} q e^{iwt} - w^2 \sum_{r=0}^{\infty} w^r a_{ry} q e^{iwt} = 0 \qquad (3)$$

and using the shape function and the algebraic manipulation presented in [14], the equation of motion for a free end – free end horizontal nonlinear microbeam (Figure 3) can be rewritten with the first order kinetic and strain terms

$$F_{iNL} = \frac{1}{2}\left(\frac{d^2q}{dt^2}^T M \frac{d^2q}{dt^2}\right) + \frac{1}{2}\left(q^T K_L q\right) + \frac{1}{2}\left(\frac{N}{L} q^T K_{NL} q\right) \quad (4)$$

where $M$ and $K_L$ are the mass and the stiffness matrices, respectively. $N = \frac{EA(L - \overset{\centerdot}{L})}{L}$ is the axial force applied on the microbeam and $K_{NL}$ is the nonlinear geometric stiffness matrix. It is demonstrated in [14] that the proposed method for modelling linear effects in MEMS valid and nonlinear problems such as the buckling of beams have also been tackled successfully.

## IV. IMPLEMENTATION OF COSSERAT THEORY INTO HAPTIC SENSING TECHNOLOGY

The algorithm structure to simulate a Cosserat microbeam in real-time in a VRE using haptic sensing technologies can be broken down into five stages. The first phase is to convert/scale up the load generated with the PHANTOM Omni to the appropriate MEMS micro-world. The second step is to measure the displacement in the reference frame. This displacement is then translated into the moving frame where different forces are calculated and summed up to obtain the total force. Afterwards, the total force is transformed back to the system matrix.

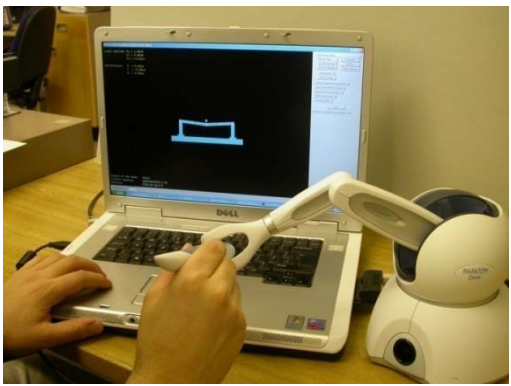

Fig. 4. Haptic interface

A scale down/conversion is subsequently carried out; the proper force feedback is then calculated and rendered. The algorithm is implemented in Object Oriented language. OpenHaptics is used to interact and to feel the MEMS components, OpenGL for graphics and Microsoft visual C++. A front end-user interface has been designed to readily interact with selected MEMS components as shown in Figure 5.

The model of a linear cantilever microbeam has then been integrated into the haptic environment as shown in figure 5. Deflection of the microbeam is occurring in real-time for a load applied by the pen of the haptic device. Figure 6 shows the deflection of another microstrcutre, a microbridge, when a load is applied with the stylus in the y-direction.

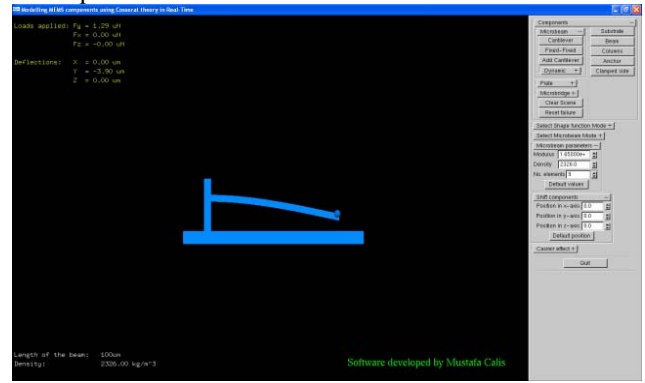

Fig. 5. Linear cantilever microbeam in deflection for a vertical load

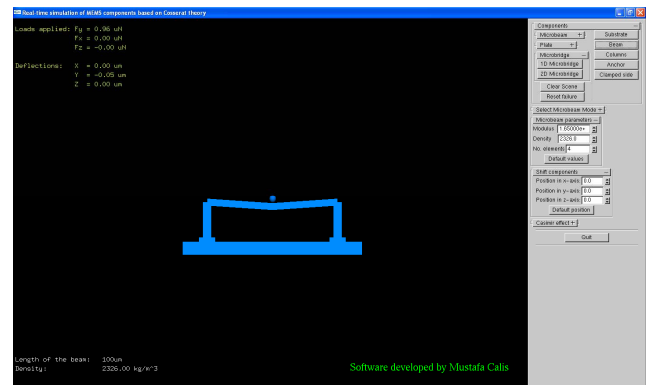

Fig. 6. Microbridge in deflection for a vertical load

Comparisons with the software packages ANSYS and SABER were not possible as these packages are not suitable for real-time simulation.

## V. MODELLING SURFACES INTERACTION

The highly intensive market for faster computing power for smaller electronic die sizes drives the development of enhanced lithographic tools for greater miniaturization. This capability is likely to be translated in the field of MEMS/NEMS. Since these devices are generally movable structures on a semiconductor and as the miniaturization is increasing, the likelihood of the moveable elements to collapse onto the substrate is augmenting concurrently (Figure 7). Therefore, for efficient manufacturing process and high performance of these devices, a thorough understanding of these surfaces interactions during growth and device operation are decisive for reliable performance. A particular category of devices falling into this category is RF-MEMS switches and relays. The effect in MEMS/NEMS is known as stiction (static friction); the adhesion of contacting surfaces which, in general results in permanent stiction. Another example of phenomenon that can cause the failure or reduce the performance of MEMS/NEMS is the Casimir force, which is of quantum nature [15]. This consists in the attraction between a pair of neutral, parallel, conductive metal plates separated by a small gap causing moveable parts to collapse into the substrate.

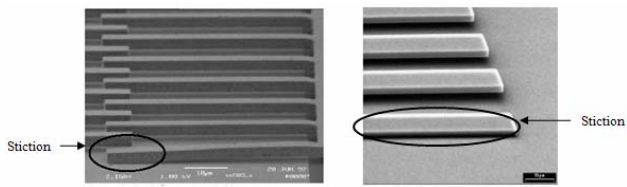

Fig. 7. Stiction effect: a) Stuck finger on comb drive-Stiction. Courtesy of Sandia National Labs, MEMS reliability departments b) Cantilever after release etch adhering ti substrate. Stiction caused by capillary forces/condensation [16].

This effect arises since the average of virtual particles is greater surrounding the plates than between them since the distance between the plates are so small that the range of fluctuations is restricted causing the generation of an attractive force (Figure 8). Casimir effect has been predicted in 1948 by the Dutch physicist Hendrik Brught Gerhard Casimir [15], but it has only been confirmed experimentally in 1997, by Steven K. Lamoreaux using a torsion balance [17].

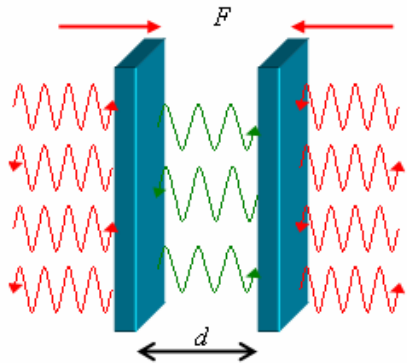

Fig. 8. Schematic representation of fluctuating virtual particles

The Casimir effect is known to be a local version of the van der Walls force between molecules and a proof for the evidence of the quantum vacuum possessing infinite energy density populated by virtual particles.

In this work, only the stiction effect which occurs when component touch the substrate is considered. The modelling of this surface interaction has been implemented in our software to guide designers during the assembly process of for training engineers quickly to operate in this field (Figure 9). With our user-end interface, the cantilever can be tested for given parameters. Moreover, the range of load that can be applied on the cantilever can be tested virtually and corrected depending on the needs at the design stage. When the cantilever is very close to the substrate, a warning messages pops-up on the screen and once it touches the substrate, the cantilever sticks to it. By pressing the "Reset Failure" on the left hand side of the window, the cantilever will be placed back in its original position. The consideration of such surfaces interaction in CAD/CAM tools can enable to build physical prototypes late in the manufacturing cycle. The advantages of using haptic sensing technologies allow the designer to touch and feel the MEMS component and at the same time to see the deflection in real-time. The parameters of the beam, such as the Young modulus, the density and number of elements composing the cantilever can be changed readily using the left hand side menu.

## VI. CONCLUSION

In this paper, it is demonstrated that Cosserat theory has been successfully integrated into haptic sensing technologies for modelling and testing simple structures such as a cantilever and a microbridge. In addition, the modelling surfaces interaction such as the stiction effect has also been integrated successfully in the software.

## REFERENCES

[1]   http://www.coventor.com

[2]   MEMSMaster of MEMSCAP Company, http://www.memscap.com

[3]   www.ansys.com

[4]   http://www.softmems.com

[5]   SABER from Synopsys, www.synopsys.com

[6]   IntelliSuite from IntelliSense, www.intellisensesoftware.com

[7]   C.B. Zilles, "Haptic Rendering with the Toolhandle Haptic Interface", Master thesis,  Massachusetts Institute of Technology, May 1995.

[8]   M. Calis, O. Laghrouche, and M.P.Y Desmulliez, "The application of Cosserat theory into haptic sensing technology: a new design methodology for MEMS", *Int. Conf. on Manufacturing Research ICMR07*, pp. 110-114, Leicester, UK, 11-13th Sept., 2007

[9]   M. Paz, and L. Dung, "Power series expansion of the general stiffness matrix for beam elements", *Intl. Journal for Numerical Methods in Engineering*, vol. 9, pp. 449-459, 1975.

[10]  M. Paz, "Mathematical observations in structural dynamics", Computers & Structures, vol. 3, pp. 385-396, 1973.

[11]  C. Wang, D. Liu, E. Rosing, B. De Masi, and A. Richardson, "Construction of nonlinear dynamic MEMS component modesl using Cosserat theory", *Analog Integrated Cicuits and Signal Processing*, vol. 40, pp. 117-130, 2004

[12]  M. Calis, O. Laghrouche, M.P.Y Desmulliez, "The implementation of Cosserat theoy into haptic sensing technology for large deflection beam model", *4th Int. Conf. on Responsive Manufacturing,* 17 – 19th September, Nottingham, UK.

[13]  J.S. Przemieniecki, "Quadratic Matrix Equations for Determining Vibration Modes", Methods Structural Mechanic *Wright-Patterson Air Force Base* 66-80, 1966.

[14]  M. Calis, O. Laghrouche, and M.P.Y Desmulliez, "A New Modelling Methodology of MEMS Structures based on Cosserat Theory", DTIP of MEMS & MOEMS, 9-11th April 2008, France.

[15]  H.B.G Casimir, Proc. Kon. Nederl. Akad. Wet B51 pp. 793, 1948

[16]  S.W. Merlijn, P. Robert, and I. De Wolf, "A Physical Model to predict stiction in MEMS", *Journal of Micromechanics and Microengineering,* vol. 12, pp. 707713

[17]  S.K. Lamoreaux, "Demonstration of the Casimir Force in the 0.6 to 6mm Range", *Phys. Rev. Lett. 78*, pp. 5-8, 1997

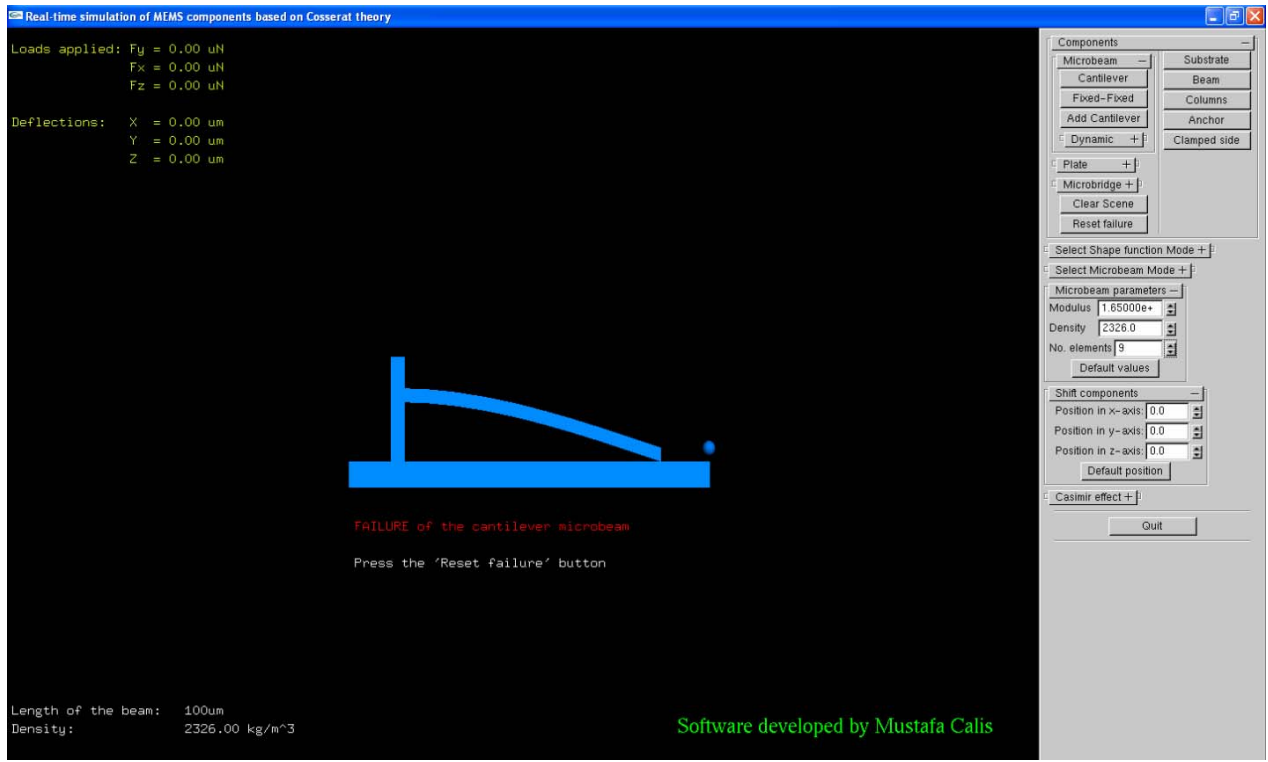

Fig. 9. Stiction effect caused by the cantilever surface touching the substrate